\documentclass[twoside]{article}

\usepackage[accepted]{aistats2022}

\usepackage[backref]{hyperref} 
\usepackage{apacite}
\usepackage{graphicx}
\usepackage{amsmath}
\usepackage{amssymb}
\usepackage{amsmath}
\usepackage{algpseudocode}
\usepackage{caption}
\usepackage{subcaption}

\begin{document}

\twocolumn[

\aistatstitle{Neural Enhanced Dynamic Message Passing}

\aistatsauthor{ Fei Gao \And Yan Zhang \And  Jiang Zhang }

\aistatsaddress{Beijing Normal University 
                \\ \href{mailto:feig@mail.bnu.edu.cn}{feig@mail.bnu.edu.cn}  
                \And Beijing Normal University 
                \\ \href{mailto:feig@mail.bnu.edu.cn}{zyan2408@mail.bnu.edu.cn} 
                \And Beijing Normal University 
                \\ \href{mailto:feig@mail.bnu.edu.cn}{zhangjiang@bnu.edu.cn}}

]

\begin{abstract}
  Predicting stochastic spreading processes on complex networks is critical in epidemic control, opinion propagation, and viral marketing. We focus on the problem of inferring the time-dependent marginal probabilities of states for each node which collectively quantifies the spreading results. Dynamic Message Passing (DMP) has been developed as an efficient inference algorithm for several spreading models, and it is asymptotically exact on locally tree-like networks. However, DMP can struggle in diffusion networks with lots of local loops. We address this limitation by using Graph Neural Networks (GNN) to learn the dependency amongst messages implicitly. Specifically, we propose a hybrid model in which the GNN module runs jointly with DMP equations. The GNN module refines the aggregated messages in DMP iterations by learning from simulation data. We demonstrate numerically that after training, our model's inference accuracy substantially outperforms DMP in conditions of various network structure and dynamics parameters. Moreover, compared to pure data-driven models, the proposed hybrid model has a better generalization ability for out-of-training cases, profiting from the explicitly utilized dynamics priors in the hybrid model. A PyTorch implementation of our model is at \href{https://github.com/FeiGSSS/NEDMP}{https://github.com/FeiGSSS/NEDMP}.
\end{abstract}

\section{Introduction}
Stochastic spreading processes on complex networks are widely used for modeling the epidemic spreading, information propagation, and transmission of social behaviors. Given the initial states, accurately predicting the spreading is of primary importance in many domains. Since the stochastic nature of the processes, the more reasonable is to predict the marginal probabilities of each state for each node, as shown in Figure \ref{fig:task}. In addition to being an accurate description of the spreading, the predicted marginal probabilities could also be used for inferring the source of spreading \cite{Zhu2016InformationSD}, maximizing the influence by selecting a fixed size of initially activated nodes \cite{Kempe2015MaximizingTS}, and determining an optimal set of nodes to immunize \cite{PastorSatorras2002ImmunizationOC}.

% 非NN的方法怎么做这个问题
% NN的方法怎么做这个问题

% 引入DMP DMP细致的好处，广泛的应用。然后介绍优势和劣势
Dynamics Message Passing (DMP) \cite{Karrer2010MessagePA,Shrestha2014AMA, Shrestha2015AMA, Lokhov2015DynamicME}, a special case of Belief Propagation (BP) on time trajectories, is an efficient algorithm  for inferring the marginal probabilities for stochastic spreading processes on graphs. For spreading process with unidirectional dynamics (e.g., SIR, SEIR), DMP is exact on tree graphs and asymptotically exact on locally tree-like graphs, and has a linear computational complexity in the number of  edges and spreading time steps. It typically yields accurate estimations on real sparse networks for a large class of spreading dynamics \cite{Lokhov2015DynamicME}. Therefore, as an analytical inference machine, it has been used for inferring the patient zero \cite{Lokhov2014InferringTO}, reconstructing the dynamics parameters \cite{Lokhov2016ReconstructingPO, Wilinski2021PredictionCentricLO}, optimal deployment of resources \cite{Lokhov2017OptimalDO}, and functional immunization of networks \cite{Li2020FunctionalIO}. 

However, same as other Belief Propagation algorithms, the fundamental assumption of DMP is the independence of neighboring messages, which limits the ability of DMP to capture high-order interdependencies, i.e., DMP can struggle in graphs with short local loops. As illustrated in Figure \ref{fig:dmp_case}, since the existence of local loops, DMP fails to approximate marginal probabilities in an example graph with only four nodes. Unfortunately, it is extremely challenging to handle message dependence in those loops analytically \cite{Cantwell2019MessagePO}.

The success of Graph Neural Networks (GNNs) in modelling complex pair-wise interactions \cite{SanchezGonzalez2020LearningTS, Fetaya2018NeuralRI, Cranmer2020DiscoveringSM, Bapst2020UnveilingTP} motivates us to model the dependence of messages in local loops using trainable GNNs. Specifically, GNNs can be used to learn a refined aggregation operation for messages beyond the naive independent assumption used in DMP. Several works have applied GNNs to improve the estimation accuracy of Belief Propagation algorithms. Gated recurrent GNNs are used in \cite{Yoon2019InferenceIP} as an end-to-end trainable inference algorithm for Binary Markov Random Fields, and learned GNNs outperforms BP in loopy graphs. More recent works focus on integrating the advantages of data-driven and model-based methods. Instead of using a scalar damping factor, which is proposed to accelerate the convergence of BP, Kuck et al.\cite{kuck2020belief} replace the constant scalar factor with a trainable Neural Networks layer. The proposed hybrid model converges much faster than using scalar factors while returning an estimate of comparable quality. Methods in \cite{Satorras2021NeuralEB, Liang2021NeuralEB} aim to improve the estimation accuracy of BP for discrete and continuous random variables, respectively. Since the inaccuracy of BP comes from the oversimplified aggregation scheme of neighboring messages,  those methods incorporate the GNNs into BP iterations as a trainable aggregation block, which is used to refine the aggregated messages in BP by learning from supervision data the complex dependence of messages. The resulting hybrid model runs BP and GNNs co-jointly, and benefits from the combination of physics prior and data-driven neural networks. However, no work has applied GNNs to improve the performance of DMP for better estimation of marginal probabilities of spreading process on graphs.

% 把贡献写成列表
Inspired by \cite{Yoon2019InferenceIP, Satorras2021NeuralEB}, in this work, instead of analytically modeling the complex dependence of messages in local loops, we use a GNNs module to learn from simulation data.  
%an we propose hybrid models to improve the performance of DMP. Firstly, by aligning the iteration scheme of DMP with recurrent graph neural networks, there are two mappings from DMP to GNNs, same as in \cite{Yoon2019InferenceIP}, i.e. GNNs running on nodes and edges (i.e. on line graph) respectively, which results in two pure data-driven models \textit{\textbf{GNN}} and \textit{\textbf{LGNN}}. 
In order to utilize the dynamical priors encoded in DMP equations, we use GNN module only to refine the inaccuracy aggregation operation in the iteration of DMP, while the exact iteration rules in DMP remain the same. The resulting hybrid model, which we call Neural Enhanced Dynamic Message Passing (NEDMP), runs DMP and GNN jointly, and benefits from complementation of model-based and data-driven components. 
For better training, we design a penalty term to enforce the monotone of predicted probabilities, which conforms the physical prior of dynamics. %
% Utilizing the popular epidemic model Susceptible-Infected-Recovered (SIR), we conduct experiments of marginal probabilities inference on various graph structures and dynamics parameters. The results shows signiﬁcantly improvement on DMP 
To verify the effectiveness of the hybrid model, utilizing the popular epidemic model Susceptible-Infected-Recovered (SIR), we conduct experiments of marginal probabilities inference on various graph structures and dynamics parameters. 
% The experiments of inference in within the training distribution validate that the proposed models perform signiﬁcantly better than DMP. And the experiments of generalization further show that the hybrid model NEDMP has generalized much better than pure data-driven models.
The main contributions can be summarized as follows: (1) We propose a customized GNNs that runs on line graph, (2) We propose a hybrid model for marginal inference, which runs GNNs and DMP jointly, (3) We design a dynamics-inspired penalty term to train the model, and (4) We conduct experiments on marginal probabilities inference problem on various graphs and dynamics parameters. The results show that all the proposed models outperform DMP, and the hybrid model, NEDMP, generalizes better than pure data-driven models. 
\section{Background}
Dynamic message passing has been derived as the inference algorithm for a wide range of diffusion processes. Without loss of generality, we use SIR model to demonstrate the spreading process on networks as well as the marginal probability inference problem.

% \subsection{Diffusion Dynamics and Marginal Probability Inference Problem}
\subsection{Problem Setting}

For the discrete SIR model on a diffusion graph $\mathcal{G}=(\mathcal{V}, \mathcal{E})$, where $\mathcal{V}$ is the set of nodes and $\mathcal{E}$ is the set of interactions, each node will take one of the three states: susceptible (S), infected (I) or recovered (R) at a specific time. Let $\sigma_i^t \in \{S, I, R\}$ be the state of node $i$ at time step $t$, the states transition follows:
\begin{equation}
 \label{SIR:def}
 \begin{split}
 & \mathbf{P}(\sigma_i^{t+1}=I | \sigma_i^t=S) = 1-\prod_{j\in \mathcal{N}_i}\left( 1 - \beta_{ji} \delta_{\sigma_j^t, I} \right) \\
 & \mathbf{P}(\sigma_i^{t+1}=R | \sigma_i^t=I) = \gamma_i,
 \end{split}
\end{equation}
where $\mathcal{N}_i$ is the set of neighbors of node $i$, $\beta_{ji}\in [0,1]$ is the infection rate of edge $(j\rightarrow i)$,  $\gamma_i \in [0,1]$ is the recovery rate of node $i$, and $\delta$ is the Kronecker function. 

Define the marginal probabilities of node $i$ as:  
\begin{equation*}
    \begin{split}
        &P_S^{i}(t)=\mathbf{P}(\sigma_i^t=S),\\
        &P_I^{i}(t)=\mathbf{P}(\sigma_i^t=I),\\
        &P_R^{i}(t)=\mathbf{P}(\sigma_i^t=R),
    \end{split}
\end{equation*}
we now formulate the problem as follows:

\textbf{Marginal Probability Inference (MPI).} For the SIR model on directed graph $\mathcal{G}=(\mathcal{V}, \mathcal{E})$ with infection rates $\{\beta_{ij}\}_{(i\rightarrow j)\in \mathcal{E}}$, recovery rates $\{\gamma_i\}_{i\in \mathcal{V}}$, and initial infectious nodes $\mathbb{S} \subseteq \mathcal{V}$, the goal is to infer the marginal probabilities conditioned on $\beta$,  $\gamma$ and  $\mathbb{S}$:
\begin{equation}
    \label{SIR:target}
    \{P_S^i(t), P_I^i(t),P_R^i(t)\}_{i\in \mathcal{V}, 1\leq t \leq T}
\end{equation}
where $T$ is a preset time step or the convergence time, and  $P_S^i(t) + P_I^i(t) + P_R^i(t) = 1$.

\begin{figure}[t]
    \centering
    \includegraphics[width=0.45\textwidth]{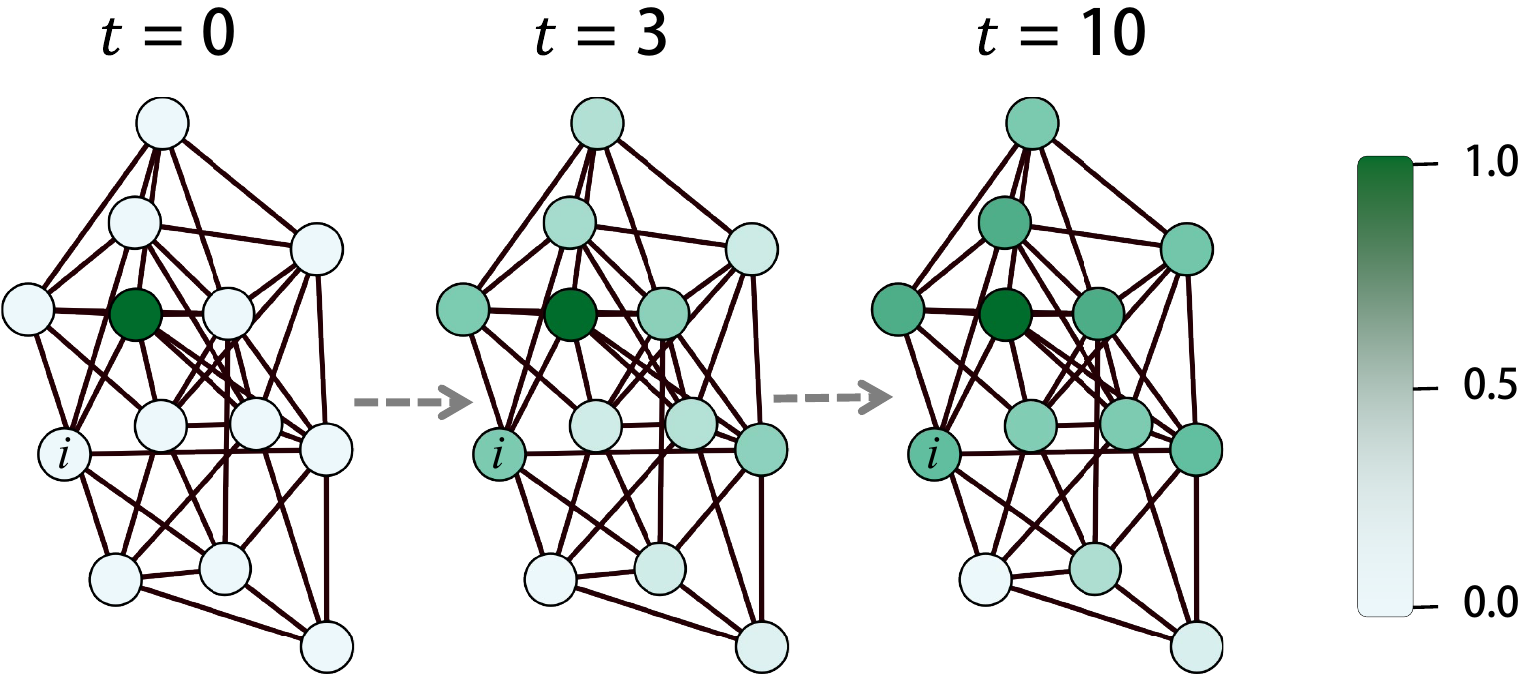}
    \caption{\textbf{Visualization of the marginal probabilities over time}, we use the SI model for illustration. Given a diffusion network with infected nodes (in darkest color) at $t=0$, the problem aims to compute $\{P_I^i(t)\}_{i\in \mathcal{V}, t\geq 1}$, i.e, the time-dependent marginal probabilities (indicated by color) of been infected for each node. The values are obtained via $10^6$ Monte Carlo simulations. }
    \label{fig:task}
\end{figure}

\begin{figure}[h]
     \centering
     
     \begin{subfigure}[b]{0.23\textwidth}
         \centering
         \includegraphics[width=\textwidth]{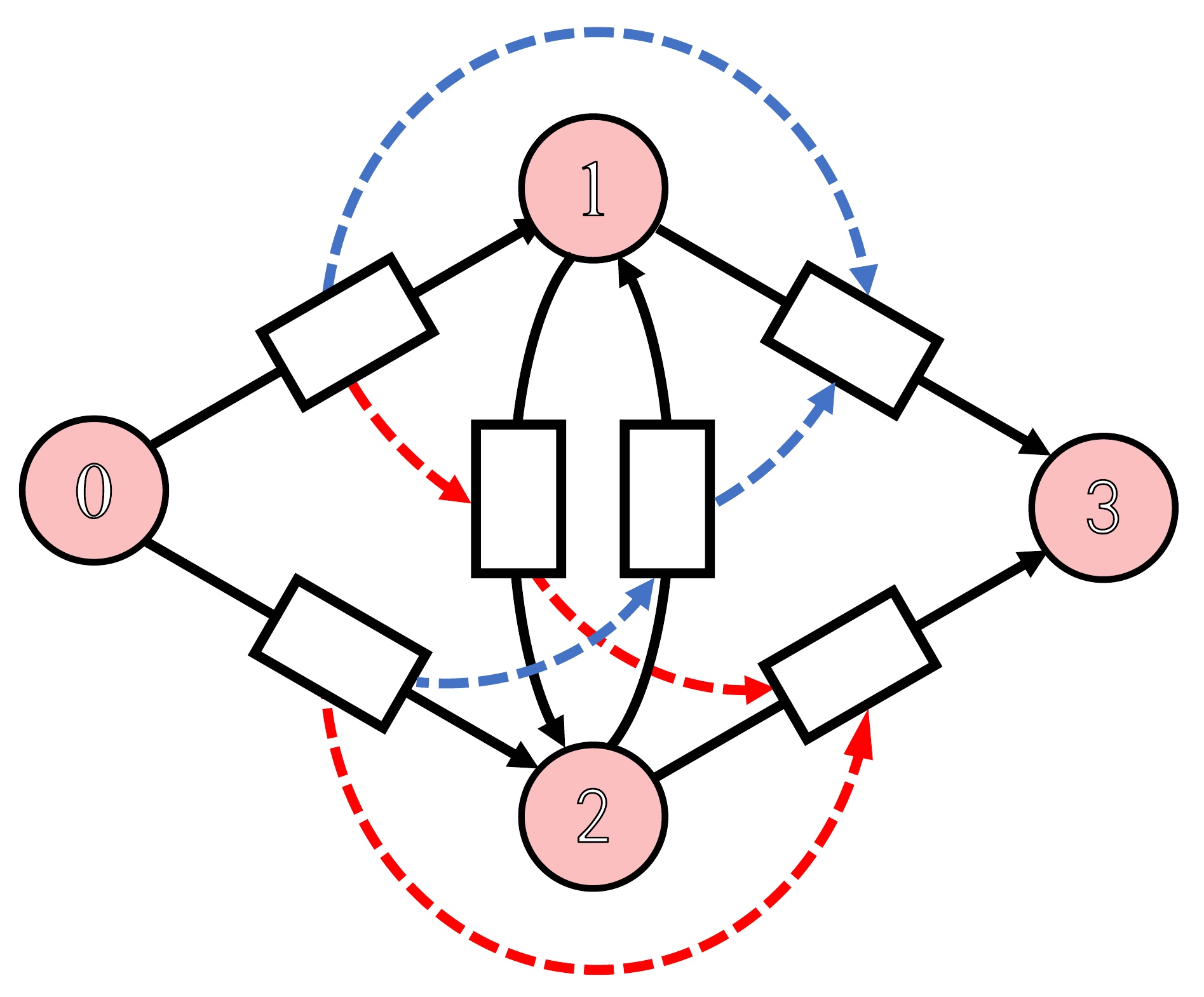}
         %\caption{An example graph. Color dashed lines indicate the dependencies of $\theta$ in DMP.}
     \end{subfigure}
     \hfill
     \begin{subfigure}[b]{0.23\textwidth}
         \centering
         \includegraphics[width=\textwidth]{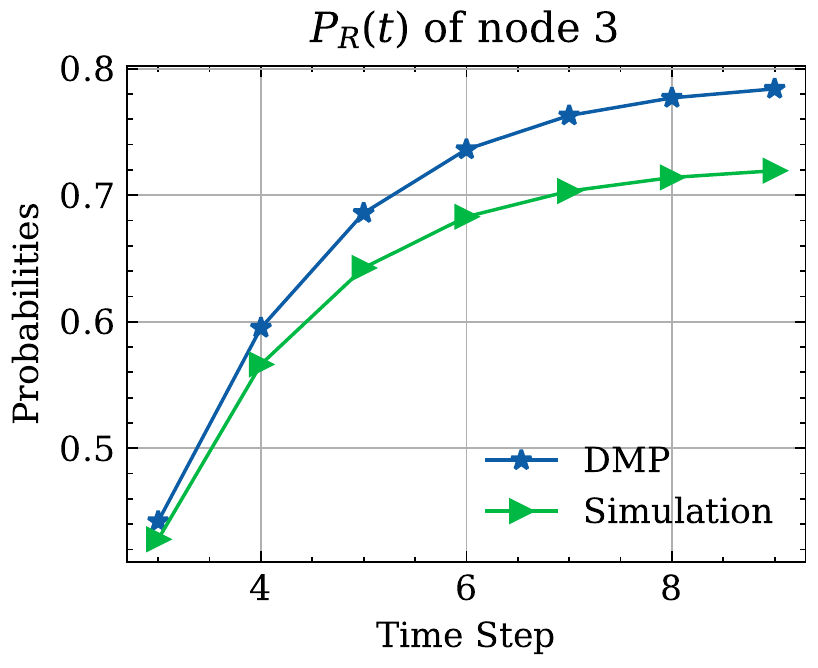}
         % \caption{The approximated marginal probability $P_R(t)$ of nodes 3 by DMP, compared to MC simulations.}
     \end{subfigure}
    \caption{\textbf{An example that illustrates the limitations of DMP.} \textit{(Left)}: A simple example graph. Color dashed lines indicate the dependencies of $\theta$ in DMP. \textit{(Right)}: The approximated marginal probability $P_R(t)$ of node 3 by DMP, compared to the \textit{true values} obtained via $10^6$ Monte Carlo simulations, where the initial conditions are $\mathbb{S}=\{0\}$, $\{\beta_{ij}\}=\{\gamma_i\}=0.5$. }
    \label{fig:dmp_case}
\end{figure}
In general, the exact computation of the marginals above is \#P-hard \cite{Wang2012ScalableIM}. Figure \ref{fig:task} is a visualization of marginals evolving over time on an example graph. We define MPI problem based on the SIR model only for convenience, and the definition can be easily generalized to other spreading models on graph, such as Independent Cascade Model \cite{Kempe2015MaximizingTS}.

\subsection{Dynamic Message Passing} \label{sec:dmp}

Dynamic message passing is the state-of-the-art method for inferring the time-dependent marginal probabilities in stochastic processes, e.g., SIR. Utilizing the unidirectionality property of the SIR model, the DMP equations for SIR are rigorously derived \cite{Lokhov2015DynamicME} from the dynamic cavity method (also known as Belief Propagation, BP) on nodes' time trajectories. It is proved \cite{lokhov2019scalable} that DMP is exact on trees and graphs without short loops. We briefly introduce the DMP equations for the SIR model here, and the detailed derivation is presented in the Appendix.

We begin with defining three intermediate dynamics variables:
\begin{itemize}
    \item $\theta^{j\rightarrow i}(t)$: the probability that disease has not spread through the edge $(j\rightarrow i)$ up to time $t$
    \item $P_S^{i\rightarrow k}(t)$: the probability that $\sigma_i^t=S$ when node $i$ ignores infection from its neighbor node $k$;
    \item $\phi^{j\rightarrow i}(t)$: the probability that disease has not spread through the edge $(j\rightarrow i)$ up to time $t$ and node $j$ is infected at time $t$ ( i.e., $\sigma_j^{t}=I$ ).
\end{itemize}

Then, the marginal probabilities of SIR model can be approximated using:
\begin{equation}
    \label{equ:pri}
    P_R^i(t) = P_R^i(t-1) + \gamma_i P_I^i(t-1),
\end{equation}
\begin{equation}
    \label{equ:pii}
    P_I^i(t) = 1 - P_R^i(t-1) - P_S^i(t-1),
\end{equation}
\begin{equation}
    \label{equ:psi}
    P_S^i(t) = P_S^i(0)\prod_{j\in \mathcal{N}_i}\theta^{j\rightarrow i}(t),
\end{equation}
\begin{equation}
    \label{equ:psik}
    P_S^{i\rightarrow k}(t) = P_S^i(0)\prod_{j\in \mathcal{N}_i\backslash k}\theta^{j\rightarrow i}(t),
\end{equation}
The value of $\theta^{j\rightarrow i}(t)$ is computed by closed iteration:
\begin{equation}
    \label{equ:theta}
    \theta^{j\rightarrow i}(t) = \theta^{j\rightarrow i}(t-1) - \beta_{ji}\phi^{j\rightarrow i}(t-1),
\end{equation}
\begin{equation}
    \label{equ:phi}
    \begin{split}
        \phi^{j\rightarrow i}(t) = 
        &(1-\beta_{ji})(1-\gamma_j)\phi^{j\rightarrow i}(t-1)\\
        & + \left(P_S^{j\rightarrow i}(t-1)-P_S^{j\rightarrow i}(t)\right),
    \end{split}
\end{equation}
with the initial values:
\begin{equation}
    \label{equ:init}
    \theta^{j\rightarrow i}(0)=1, \phi^{j\rightarrow i}(0)=\delta_{\sigma_j^0, I}
\end{equation}

% \begin{figure}[t]
%     \centering
%     \includegraphics[width=0.45\textwidth]{figs/dmp_illu_cropped.pdf}
%     \caption{\textbf{Illustration of the updating-flow of DMP equations for SIR model on a simple directed graph.} Circles and rectangles represent graph vertices and message nodes respectively. At each time step, message nodes (e.g. $i\rightarrow j$) update the corresponding dynamic variables ($\theta^{i\rightarrow j}, \phi^{i\rightarrow j}, P_S^{i\rightarrow j}$) by their former (red dashed line) and in-neighbors (blue dashed line) information. The inference of marginals is computed by a read-out aggregation of the message nodes (green dashed line)}
%     \label{fig:dmp}
% \end{figure}

Following the equations above, we can recursively compute the marginal probabilities of SIR model. % as shown in Algorithm \ref{alg:dmp}.

% As visualized in Fig. \ref{fig:dmp}, by defining each directed edge as the message node, the DMP equations can be mapped into two parts: (1) the recursive updating-flow on the line graph of message nodes adjacency (red and blue dashed lines); (2) and the read-out aggregation from message nodes to graph vertices (green dashed lines). On general graphs, the inaccuracy of DMP equations comes from the aggregation operators (Eq.(\ref{equ:psi})(\ref{equ:psik})) in those two parts. The updating scheme of DMP reminds us of the recurrent graph neural networks, which recursively update nodes embedding by non-linearly aggregating their former and in-neighbours information. The analogy of GNN and DMP inspires us to parameterize DMP with GNN and then improve the performance of DMP by supervised learning. Details are presented at the next section.

Like other BP algorithms, DMP builds on the assumption that the graph structure is a tree, i.e., messages from different neighbors are independent of each other. This allows DMP to take the simplest approach (i.e., $\mathbf{\prod}$) to aggregate the messages from neighbors, as in Equation (\ref{equ:psi}) and (\ref{equ:psik}).

However, the assumption also prevents DMP from handling the high-order dependencies between neighboring messages introduced by the local loops. As illustrated in Figure \ref{fig:dmp_case}, when executing DMP in the simple graph, $\theta^{1\rightarrow 3}(t)$ and  $\theta^{2\rightarrow 3}(t)$ are no longer independent when $t\geq 2$, since they all root from $\theta^{0\rightarrow 1}(t)$ and  $\theta^{0\rightarrow 2}(t)$ through blue and red paths respectively. Thus, the resulting marginal probabilities, when estimated using Equations (\ref{equ:psi}) and (\ref{equ:psik}),  are no longer accurate, i.e., $P_S^i(t) \neq P_S^i(0)\prod_{j\in \mathcal{N}_i}\theta^{j\rightarrow i}(t)$ and $P_S^{i\rightarrow k}(t) \neq P_S^i(0)\prod_{j\in \mathcal{N}_i\backslash k}\theta^{j\rightarrow i}(t)$.

Analytically extending DMP to high-order neighbor-hood structures is very challenging \cite{Cantwell2019MessagePO}. Therefore, a reasonable choice is to use a data-driven approach to learn a more appropriate message aggregation function that can capture the complex dependencies between messages.

\section{Method}
This section introduces our hybrid model, in which the DMP iterates jointly with a GNNs module. The GNNs module is designed to refine the aggregated messages in the iteration process of DMP.

\subsection{Graph Neural Networks on Line Graph}\label{sec:gnn}
Notice that DMP for SIR is essentially a message-passing process on line graph $\mathcal{L} = \mathcal{(N_L, E_L)}$ with non-backtracking adjacency, i.e., $\mathcal{N_L}=\mathcal{E}$ and $\mathcal{E_L}=\{(i\rightarrow j) \rightarrow (j\rightarrow k)\}_{i,j,k\in \mathcal{V}, i\neq k}$. To enable better integration of DMP and GNNs, we first define a special case of GNNs on line graph $\mathcal{L}$. Specifically, we customize and extend Gated Graph Neural Networks \cite{Li2016GatedGS} on graph $\mathcal{L}$. 

Mathematically, at every time step $t>0$, each node $(i\rightarrow j)$ in graph $\mathcal{L}$ is associated with a hidden state $h^{i\rightarrow j}(t) \in \mathbb{R}^D$. In our scenarios, node $(i\rightarrow j)$ receives time-varying inputs $x^{i\rightarrow j}(t)\in \mathbb{R}^F$. We combine the inputs with hidden states as the messages $\tilde{h}^{i\rightarrow j}(t)$:
\begin{equation}
    \tilde{h}^{i\rightarrow j}(t) = \phi_m \left(h^{i\rightarrow j}(t) \oplus \phi_e \left(x^{i\rightarrow j}(t)\right)\right),
\end{equation}
where the $\oplus$ is the concatenation operator. We then aggregate the incoming messages for target node $(i\rightarrow j)$:
\begin{equation}
    \tilde{h}^{\rightarrow (i\rightarrow j)}(t) = \phi_a (\sum_{k\neq j, (k\rightarrow i) \in \mathcal{N_L}} \tilde{h}^{k\rightarrow i}(t)),
\end{equation}
Finally, we update the hidden states for every node $(i\rightarrow j)$ based on the aggregated messages and current states via  gated recurrent unit (GRU):
\begin{equation}
    h^{i\rightarrow j}(t+1) = GRU\left(\tilde{h}^{\rightarrow (i\rightarrow j)}(t), h^{i\rightarrow j}(t)\right).
\end{equation}

The functions $\phi_m$, $\phi_e$ and $\phi_a$ are nonlinear functions mapping input to $\mathbb{R}^D$, and are instantiated as multilayer perceptron (MLP) with Rectified Linear Unit (ReLU) as the activation function. The parameters of $\phi_m$, $\phi_e$, $\phi_a$ and $GRU$ are shared by all nodes $(i\rightarrow j)$ and time steps $t$.

The equations above define one iteration of GNNs from time step $t$ to $t+1$ on line graph $\mathcal{L}$. To achieve time-dependent prediction, we can iterate those equations and  feed $\{\tilde{h}^{i\rightarrow j}(t)\}_{t>0}$ to a nonlinear readout function $\phi_{\mathcal{R}}$:
\begin{equation}
    \hat{y}^i(t) = \phi_{\mathcal{R}}(\sum_{k, (k\rightarrow i) \in \mathcal{N_L}} \tilde{h}^{k\rightarrow i}(t)),
\end{equation}
\begin{equation}
    \hat{y}^{i\rightarrow j}(t) = \phi_{\mathcal{R}}(\tilde{h}^{\rightarrow (i\rightarrow j)}(t)),
\end{equation}
where $\hat{y}^i(t)$ and $\hat{y}^{i\rightarrow j}(t)$ corresponding to the node-wise and edge-wise prediction in graph $\mathcal{G}$.

\subsection{Neural Enhanced Dynamic Message Passing (NEDMP)} \label{sec:nedmp}

As discussed in Section (\ref{sec:dmp}), the inference inaccuracy of DMP is due to its inability to capture the high-order interdependencies while the graph neural networks are good at capturing high-order relations \cite{Wu2019ACS}. In order to break this limitation of DMP and also to preserve the precise physically inspired part of DMP, we propose a hybrid model of GNNs and DMP, which is an extension of NEBP in \cite{Satorras2021NeuralEB}. 

Specifically, the hybrid model runs GNNs and DMP on the line graph $\mathcal{L}$ jointly. The GNNs module preserves and updates the hidden states for each node by incorporating the messages from DMP iterations; in return, the GNNs module outputs a refinement of aggregated messages in DMP. We name this model Neural Enhanced Dynamic Message Passing (NEDMP), which benefits from the complementary strengths. 
\begin{figure}[t]
    \centering
    \includegraphics[width=0.45\textwidth]{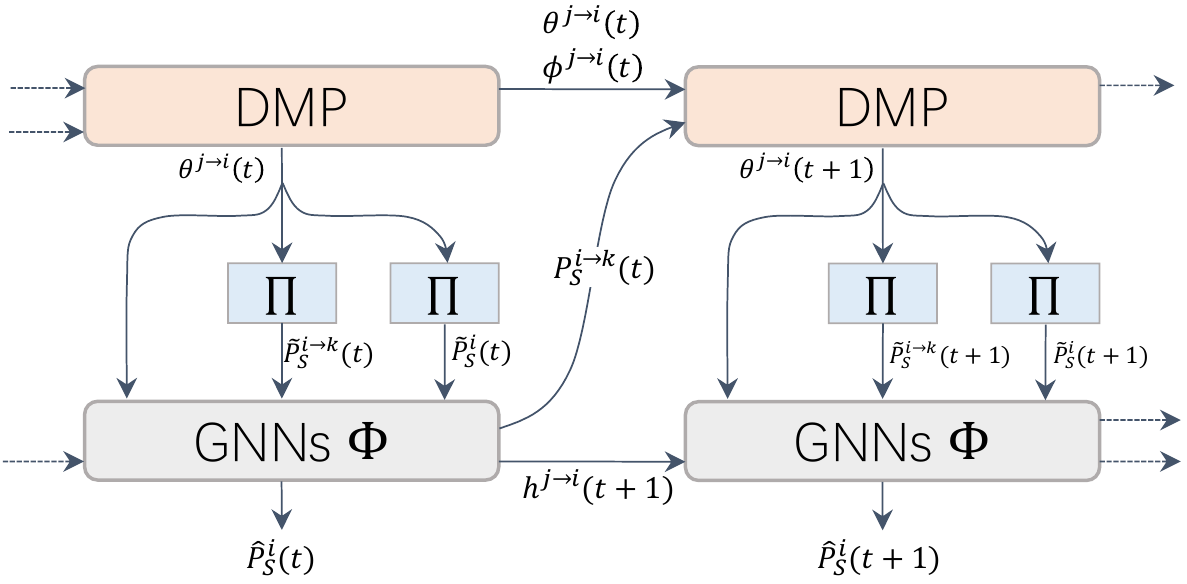}
    \caption{\textbf{Visualization of NEDMP}.} 
    \label{fig:nedmp}
\end{figure}

We adopt the GNNs variant introduced in Section (\ref{sec:gnn}), termed as $\Phi$. And the messages $\{\theta^{i\rightarrow j}(t)\}$ from DMP are provided as the inputs $\{x^{i\rightarrow j}(t)\}$ to $\Phi$. We initialize $\{h^{i\rightarrow j}(0)\}$ as:
\begin{equation}
    h^{i\rightarrow j}(0) = \phi_e(\theta^{i\rightarrow j}(0))
\end{equation}

Since we aim to refine the aggregated messages $\tilde{P}_S^i(t)$ and  $\tilde{P}_S^{i\rightarrow k}(t)$ in DMP iteration, it is reasonable to integrate those values into $\Phi$ as the prediction baseline. Therefore, we modify the readout function in $\Phi$ as follows:
\begin{equation}\label{equ:refi1}
    \begin{split}
        &\xi^{i}(t), \zeta^{i}(t) =\\
        &\phi_{\mathcal{R}}\left(\tilde{P}_S^i(t) \oplus \sum_{k, (k\rightarrow i) \in \mathcal{N_L}} \tilde{h}^{k\rightarrow i}(t)\right),
    \end{split}
\end{equation}
\begin{equation}\label{equ:refi2}
    \begin{split}
        & \xi^{i\rightarrow j}(t), \zeta^{i\rightarrow j}(t) = \\
        & \phi_{\mathcal{R}}\left(\tilde{P}_S^{i\rightarrow j}(t) \oplus \tilde{h}^{\rightarrow (i\rightarrow j)}(t)\right)
    \end{split}
\end{equation}
where $\phi_{\mathcal{R}}$ is a nonlinear function mapping inputs to $[0,1]^2$, and is instantiated as multilayer perceptron (MLP) with Sigmoid as the activation function. Utilizing the readouts of module $\Phi$, we can finally refine the messages by applying  affine transformation:
\begin{equation}\label{equ:refin_psi}
    P_S^{i}(t) = \tilde{P}_S^{i}(t)\cdot \xi^{i}(t) + \zeta^{i}(t),
\end{equation}
\begin{equation}\label{equ:refin_psik}
    P_S^{i\rightarrow k}(t) = \tilde{P}_S^{i\rightarrow k}(t)\cdot\xi^{i\rightarrow k}(t) + \zeta^{i\rightarrow k}(t), 
\end{equation}
The refined $P_S^{i}(t)$ and $P_S^{i\rightarrow k}(t)$ are fed back to DMP for the next iteration. The whole framework of NEDMP is visualized in Figure \ref{fig:nedmp}. % We formulate the inference procedure as Algorithm \ref{alg:nedmp}.

\subsection{Training}
% Since the convergence time step of marginals varies from different graphs and different dynamics variables, it is improper to pre-define a fixed number of iteration of the proposed models. Therefore, we add a breaking mechanism to the models to enable them to jump out of iteration adaptively. Specifically, the model will monitor the changes of marginals between two consecutive layers:
% \begin{equation}
%     \Delta = \max{|\hat{P}(t)-\hat{P}(t-1)|}
% \end{equation}
% and the models break the iteration when $\Delta < 5\times 10^{-3}$. 
We can obtain the \textit{ground-truth} marginal probabilities $q^i(t) = [q_S^i(t), q_I^i(t), q_R^i(t)]$ by extensive Monte Carlo simulations. Therefore, the simplest way to optimize the proposed model NEDMP is to minimize the cross-entropy($CE$) loss:
\begin{equation}
    \label{equ:los1}
    l(P, q) = \frac{1}{|\mathcal{V}|T}\sum_{i\in \mathcal{V}} \sum_{1\leq t \leq T} CE(P^i(t), q^i(t))
\end{equation}

% TODO hat是生成的
where $T$ is the preset time length or the convergence time step, $P^i(t)$ is the predicted probabilities by model. It is worth noticing that the dynamics of SIR model ensure the monotone of node's marginals, i.e., $\forall t_1 \leq t_2, i\in \mathcal{V}$: $P_S^i(t_1) \geq P_S^i(t_2)$ and $P_R^i(t_1) \leq P_R^i(t_2)$, which the sole objective in Equation (\ref{equ:los1}) may fail to capture. Thus, we add a regularization term to enforce this limitation:
\begin{equation}
\begin{split}
    & Re =  \\
    & \sum_{i\in \mathcal{V}} \sum_{1\leq t \leq T} [ReLu(P_S^i(t+1) - P_S^i(t)) \\
    & +relu(P_R^i(t) - P_R^i(t+1))]
\end{split}
\end{equation}
We combine the regularization to the cross-entropy loss with a factor $\lambda$ as the final objective function:
\begin{equation}
    \label{equ:loss}
    L(P, q) = l(P, q) + \lambda\cdot Re
\end{equation}
\section{Experiments and Results}

\begin{figure*}[t]
\centering
\includegraphics[width=0.7\textwidth]{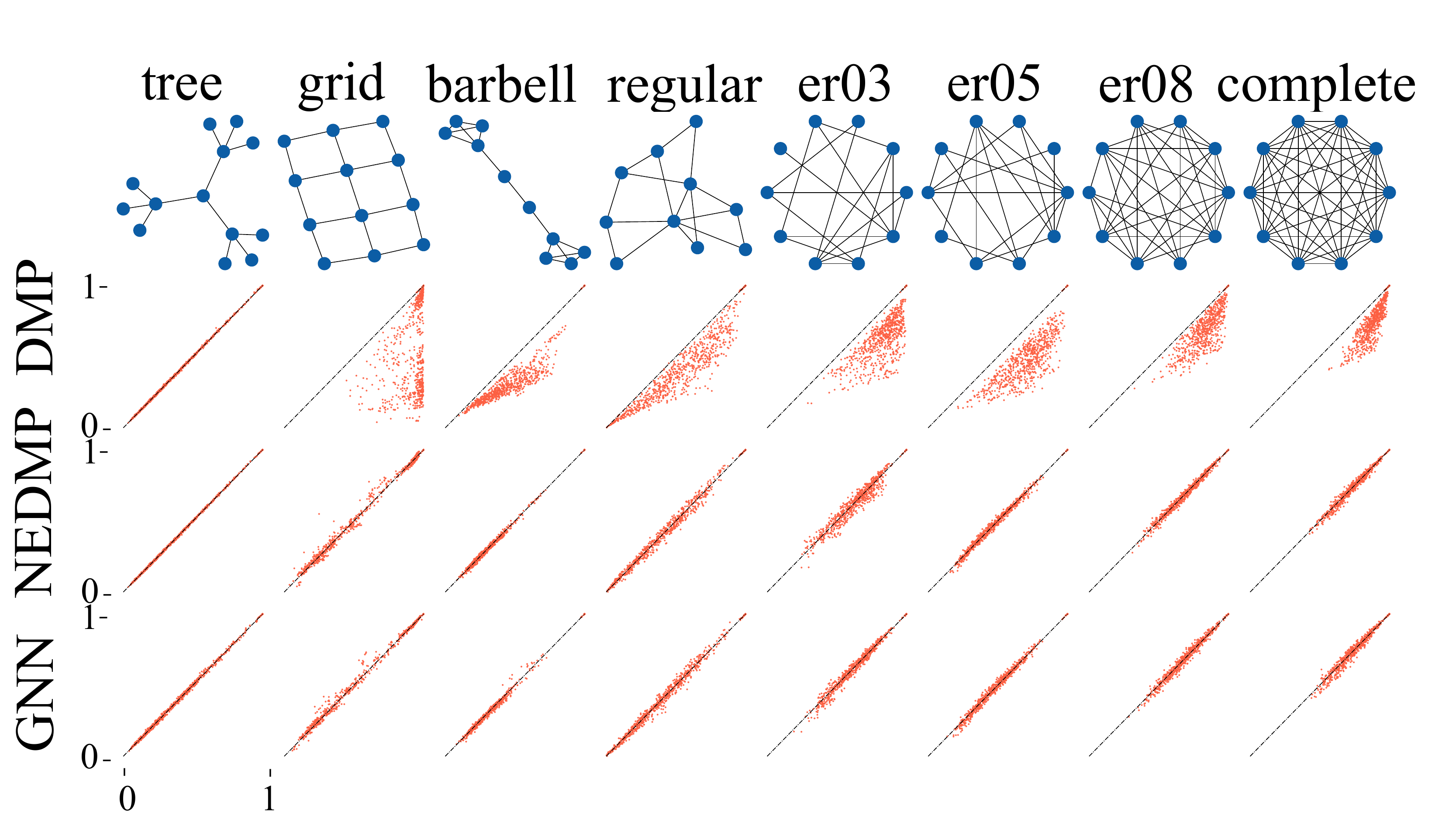}
\caption{\textbf{Performance on diverse graph structure.} For each graph structure, the estimated marginals are plotted as horizontal coordinates while the vertical coordinates represent the ground truth. For visual simplicity, we only show the marginal probability $P_R^i(T)$ for each node $i$ in testing graphs. The corresponding dot will lie on the diagonal line if it is accurately estimated.} 
\label{fig:syn_visu}
\end{figure*}

We conduct experiments on a set of synthetic and real networks to evaluate the performance of the proposed model NEDMP with respect to the MPI problem of the SIR model. Specifically, in Section \ref{sec:performance} we evaluate the performance of all methods on diverse sets of graphs and dynamics parameters, where the training and testing sets come from the same distribution. In Section \ref{sec:generalization}, we evaluate the generalization ability of the proposed models on instances that have graph structure or dynamics parameters out of the training distribution.

\textbf{Data Generation.} For the training and testing in all experiments, we obtain the \textit{ground-truth} of marginal probabilities by averaging over $10^5$ Monte Carlo simulations.

\textbf{Baselines. } We compare three methods for the MPI problem of SIR model:
\begin{itemize}
    \item \textbf{DMP} \cite{Lokhov2014InferringTO}: An efficient inference algorithm for SIR model, and it is asymptotically exact on locally tree-like networks.
    \item \textbf{GNN} \cite{Yoon2019InferenceIP}: This work proposed two pure data-driven models for inferring the marginal probabilities for probabilistic graphical models. Considering the variant msg-GNN (similar to the model proposed in Section (\ref{sec:gnn})) increases the computational consumption, with no improvement in accuracy, we adapt the variant node-GNN for our problem (details in Appendix).  
    \item \textbf{NEDMP}: A hybrid model in which DMP and GNNs module runs jointly. The GNNs module is designed to improve the accuracy of messages in DMP by learning the high-order dependencies from simulation data. 
\end{itemize}

\textbf{Evaluation Metrics.} Averaged L1 error is used as the performance metrics for all experiments: 
\begin{equation}
    L_{1} = \frac{1}{|\mathcal{N}|T}\sum_{i\in \mathcal{N}} \sum_{1\leq t \leq T} \lVert P^i(t) - q^i(t) \rVert_1
\end{equation}

\textbf{Training Details.} We optimize the proposed models with loss function in Eq.(\ref{equ:loss}) with $\lambda=5$. The maximum time step $T$ is 30. We use the Adam optimizer\cite{Kingma2015AdamAM} with the learning rate $0.01$ and batch size of $1$. The learning rate reduces by a factor of $0.5$ whenever the learning stagnates. We use early stopping with patience $15$. The dimensions of all hidden states are $D=32$. Each dataset in the following experiments is split by 6:2:2 for training, validation, and testing, respectively.

\subsection{Performance Within the Training Distribution } \label{sec:performance}

The complexity of MPI comes from two parts: the topology structure of the graph as well as the values of dynamics parameters, i.e., $\beta$, $\gamma$ and $\mathbb{S}$. Therefore, it is necessary to train and test the methods over different graph structures and parameter ranges. Particularly, the training and testing set are generated within the same distribution of $\beta$, $\gamma$ and $\mathbb{S}$. We also evaluate the methods on several real networks. 

\begin{figure*}[t]
\centering
\includegraphics[width=0.9\textwidth]{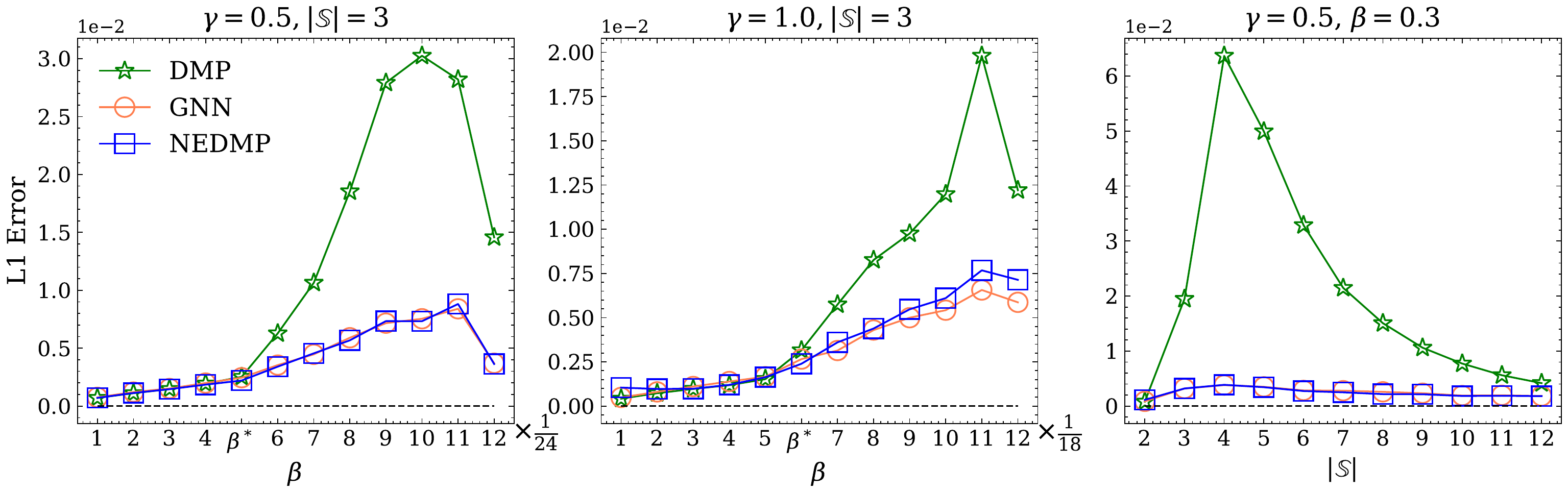}
\caption{\textbf{Performance of various dynamics parameters.} In the \textbf{left} and \textbf{middle} figures, the graph structures are fixed to a 3-regular graph with $|\mathcal{N}| = 20$, and we increase the graph size to $|\mathcal{N}| = 100$ in the \textbf{right} figure. $\beta^*$ in the \textbf{left} and \textbf{middle} figure is the tipping point of diffusion.} 
\label{fig:performance_parameters}
\end{figure*}  

\begin{table*}[h]
    \caption{\textbf{$\mathbf{L_1}$ error on real networks (lower is better).}}
    \centering
    \begin{tabular}{ccccccc}
    \hline
      & dolphins       & fb-food        & fb-social      & norwegain      & openflights    & top-500\\ 
    \hline
    $\#$Nodes   & 62         & 620      & 1899      & 1482           & 2939           & 500\\
    $\#$Edges   & 159        & 2102     & 20296     & 4006           & 15677          & 2980\\
    \hline
    DMP    & 0.089 $\pm$ 0.126  & 0.096 $\pm$ 0.177     & 0.023 $\pm$ 0.062     & 0.103 $\pm$ 0.199     & 0.077 $\pm$ 0.125     & 0.064 $\pm$ 0.089\\ 
    GNN    & 0.028 $\pm$ 0.033  & 0.030 $\pm$ 0.039     & 0.021 $\pm$ 0.037     & 0.018 $\pm$ 0.042     & 0.032 $\pm$ 0.044     & 0.033 $\pm$ 0.048\\ 
    NEDMP  & 0.027 $\pm$ 0.036  & 0.034 $\pm$ 0.044     & 0.024 $\pm$ 0.043     & 0.023 $\pm$ 0.054     & 0.048 $\pm$ 0.064     & 0.032 $\pm$ 0.044\\
    \hline
    \end{tabular}
    \label{tab:real_res}
\end{table*}

\textbf{Graph Structures.} As shown in Figure \ref{fig:syn_visu}, we choose eight types of graph structure with the number of nodes $|\mathcal{N}|  \approx 12$. For each structure, we generate 200 samples for training and testing. Each of the samples has one randomly chosen node as the initial infectious seed, and dynamics parameters from $\beta \sim \mathcal{U}(0.4, 0.6)$ and $\gamma \sim \mathcal{U}(0.2, 0.5)$. After training, we run each method on testing graphs, and plot the inferred marginal probabilities, paired with ground truth, at time step $T$ for each method in Figure \ref{fig:syn_visu}. The first row in Figure \ref{fig:syn_visu} verifies the inaccuracy of DMP in loopy graphs, as well as that DMP, is the upper bound of true influence \cite{lokhov2019scalable}. The flexibility and powerful inference ability of both GNN and NEDMP are validated by the bottom two rows which show almost accurate inferences in all kinds of graphs.

\textbf{Dynamics Parameters.} With the structure fixed as the n-regular graph, we vary the dynamics parameters $\beta$, $\gamma$, and $\mathbb{S}$ around the tipping point to evaluate the performance of proposed models, as shown in Figure \ref{fig:performance_parameters}. For each combination of parameters, we generate 200 samples for training and testing. When parameters approach the poles of the x-axis,  the diffusion steps are too few to form loopy paths, resulting in trivial instances for MPI problem. This is why the error curve of DMP is bell-shaped. The two trained models GNN and NEDMP have similar performance, and both outperform DMP in all ranges of parameters. As parameters approach the area with loopier diffusion paths, the performances of the learning-based models degrade much slower than DMP.

\textbf{Real Diffusion Networks.} We also train and test all the methods on six true diffusion networks  \cite{nr-aaai15}, and all the graphs are preprocessed as undirected graphs. Again, we generate 200 samples for each graph as the training and testing set. For each sample, two randomly selected nodes are set as the initially infected nodes. And dynamics parameters are randomly sampled from $\beta \sim \mathcal{U}(0, 0.3)$ and $\gamma \sim \mathcal{U}(0.1, 0.4)$. Same as previous, within the distribution of training set, GNN and NEDMP have comparable performances, and all outperform DMP based on metrics $L_1$.

\begin{figure*}[t]
\centering
\includegraphics[width=0.9\textwidth]{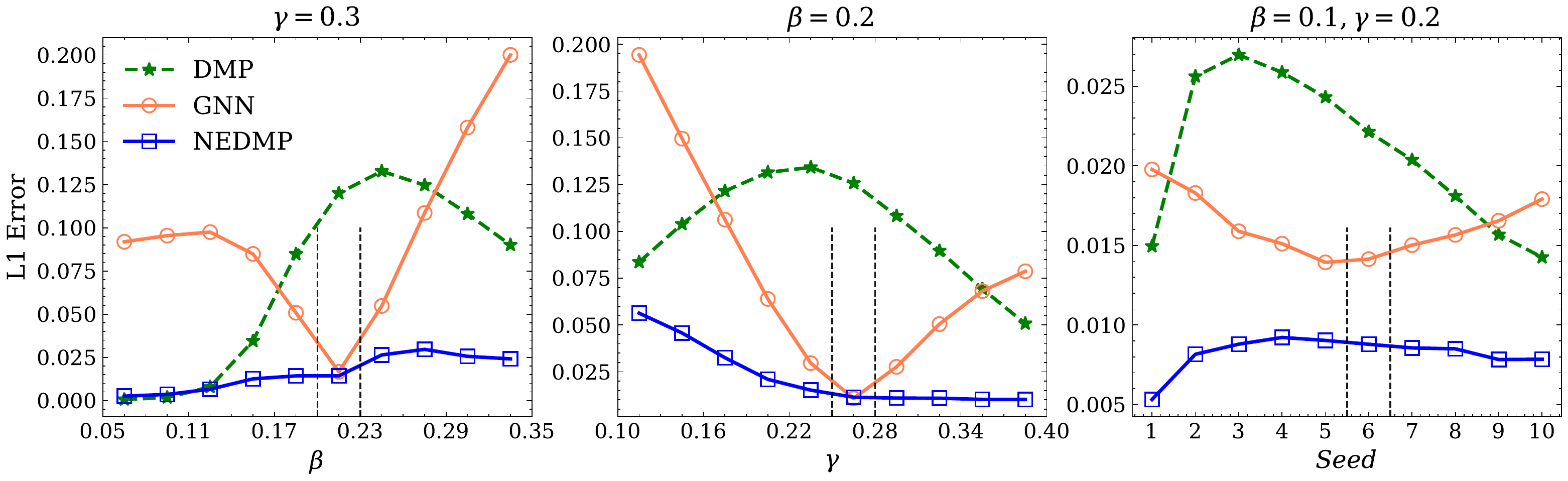}
\caption{\textbf{Generalization Performance on Dynamics Parameters.} GNN and NEDMP are trained in a small range of parameter values, bounded by two vertical dashed lines, and then tested out-of-set. The graph structure is fixed to be Watts–Strogatz graph with 50 nodes, each node connects 5 nearest neighbors and the probability of rewiring each edge is $p=0.2$.} 
\label{fig:gene_parameter}
\end{figure*} 

\begin{figure*}[t]
    \centering
    \includegraphics[width=0.9\textwidth]{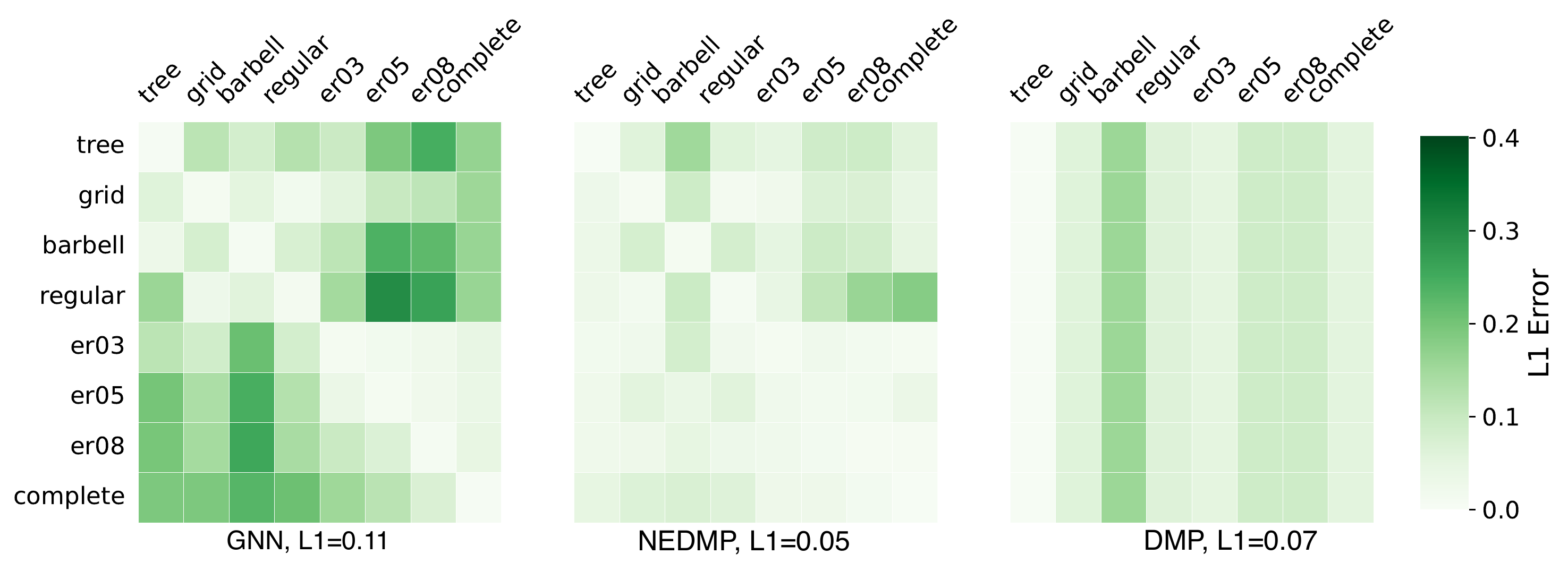}
    \caption{\textbf{Generalization on Graph Structure.} (a,b): GNN and NEDMP are trained on a specific graph structure and then test on all other graph structures. The diagonals are the training errors while the off-diagonal cells represent the generalization errors for each model. The average of all off-diagonal cells is labelled at the bottom of each heatmap.} 
\label{fig:gene_structure}
\end{figure*}

\subsection{Out of Distribution Generalization} \label{sec:generalization}
It is well known that the neural network model suffers from poor generalization. Using \textit{hand-engineering} and \textit{end-to-end} learning cooperatively, a hybrid model which benefits from their complementary strengths can be one way to break the generalization limitation \cite{Shlezinger2020ModelBasedDL}. The proposed NEDMP is such a hybrid model while GNN is a purely data-driven model. To better understand the generalization ability of those two kinds of models, we conduct experiments on structure and parameter generalization.

\textbf{Structure Generalization.} We freeze the models trained on specific graph structures in Section \ref{sec:performance}, and then test them on all eight graph structure datasets. For instance, a GNN model trained on tree graphs is then tested on all graph structures, and the eight testing $L_1$ errors are visualized as the first row in Figure \ref{fig:gene_structure}(a). All the off-diagonal cells in Figure \ref{fig:gene_structure}(a,b) are the generalization errors for the corresponding model. The average of generalization errors are 0.128, 0.035 for GNN and NEDMP respectively. The superiority of NEDMP in generalization verifies the strengths of the hybrid model.

\textbf{Dynamics Parameters Generalization. } We consider the model generalization over $\beta$, $\gamma$ and $\mathbb{S}$. We fix the graph structure as the Watts–Strogatz graph. As shown in Figure \ref{fig:gene_parameter}, for each parameter spectrum, we train the GNN and NEDMP on a small range of parameter values (bounded by the two vertical dashed lines). Then the trained models are tested on the whole parameter spectrum; the out-of-set results are presented in Figure \ref{fig:gene_parameter}. As expected, GNN degrades rapidly as the parameter away from the training set. Surprisingly, NEDMP can maintain almost the same performance outside the training distribution, and it even outperforms DMP under all unseen parameters. NEDMP is so robust to parameters because the GNN module in NEDMP does not interact with parameters directly; parameters are encoded by well-established updating equations in the DMP module, which enables the hybrid model to generalize to unseen values.

\section{Discussion}
In this work, we propose a hybrid model Neural Enhanced Dynamic Message Passing (NEDMP), which runs DMP and GNN jointly, for the marginal probabilities inference problem of the SIR model. We test the performance of inference in diverse sets of graph structures and dynamics parameters, and the results show that after training the proposed model significantly outperforms DMP in all kinds of graph structure  within the training distribution. The generalization ability of the proposed model is evaluated by inferring on graph structures and parameters that are out of the training distribution. The results show that NEDMP generalizes much better than the pure data-driven model since it incorporates the informative dynamics-based prior bias from DMP module. 

This work is also a demonstration of the emerging field of hybrid physics-guided machine learning \cite{Rai2020DrivenBD}. We use physics priors (DMP) to guide the design of the neural networks and regularize the training process. In future work, incorporating the physic prior and graph neural networks more compactly can be a promising direction.

As demonstrated in the experiments, NEDMP can be an efficient estimator for the spreading processing on graphs, which makes it potentially useful for forecasting and controlling epidemics (e.g., COVID-19) or rumor spreading on social networks. However, since NEDMP relies on the line graph of spreading networks, it is computationally hard to be applied to large social networks. Reducing the complexity of NEDMP is requisite for realistic scenarios, which is left for future work.
\section*{Acknowledgements}
We are grateful for supporting from "Save 2050 Project" which is sponsored by Swarma Club and X-Order, and we also thank Muyun Mou and Jing Liu for technics supporting and insightful discussion.
\bibliography{references}

%%%%%%%%%%%%%%%%%%%%%%%%%%%%%%%%%%%
%%%%%% SUPPLEMENT (OPTIONAL) %%%%%%
%%%%%%%%%%%%%%%%%%%%%%%%%%%%%%%%%%%

\clearpage
\appendix
\thispagestyle{empty}

% For one-column format, uncomment the following:
\onecolumn \makesupplementtitle
% For two-column format, uncomment the following:
%\twocolumn[ \makesupplementtitle ]

\section{Derivation of DMP Equations for SIR Model}

We follow \cite{Lokhov2015DynamicME} to introduce the DMP equations for the SIR model as well as the physical sense of intermediate dynamic variables.

We start with the updated rules of $P_R^i(t)$ and $P_I^i(t)$, derived directly from SIR model:
\begin{equation}
    P_R^i(t) = P_R^i(t-1) + \gamma_i P_I^i(t-1)
\end{equation}
\begin{equation}
    P_I^i(t) = 1 - P_R^i(t-1) - P_S^i(t-1) 
\end{equation}
Let $\theta^{j\rightarrow i}(t)$ be \textit{the probability that disease has not spread through the edge $(j\rightarrow i)$ up to time $t$}, and $P_S^i(t)$ is updated by Eq.(\ref{equ:psi}) which is exact on the tree and approximate on general graphs :
\begin{equation}\label{equ:psi}
    P_S^i(t) = P_S^i(0)\prod_{j\in \mathcal{N}_i}\theta^{j\rightarrow i}(t)
\end{equation}
Before deriving the close form of $\theta^{j\rightarrow i}(t)$, we introduce two useful intermediate dynamic variables: 
\begin{itemize}
    \item $P_S^{i\backslash k}(t)$: \textit{the probability that $\sigma_i^t=S$ when node $i$ ignores all infection from its neighbor node $k$};
    \item $\phi^{j\rightarrow i}(t)$: \textit{the probability that disease has not spread through the edge $(j\rightarrow i)$ up to time $t$ and node $j$ is infected at time $t$ ( i.e., $\sigma_j^{t}=I$ )}.
\end{itemize}
By excluding node $k$ from $\mathcal{N}_i$ in Eq.(\ref{equ:psi}), we have
\begin{equation}\label{equ:psik}
    P_S^{i\backslash k}(t) = P_S^i(0)\prod_{j\in \mathcal{N}_i\backslash k}\theta^{j\rightarrow i}(t)
\end{equation}

Utilizing variable $\phi^{j\rightarrow i}(t-1)$, we have the update rule for $\theta^{j\rightarrow i}(t)$:
\begin{equation}
    \theta^{j\rightarrow i}(t) = \theta^{j\rightarrow i}(t-1) - \beta_{ji}\phi^{j\rightarrow i}(t-1)
\end{equation}
The change of $\phi^{j\rightarrow i}(t-1)$ comes from two parts: (1) When node $j$ is infected at time $t-1$, it becomes recovery with probability $\gamma_j$ or infects node $i$ with rate $\beta_{ji}$ in the next time step; (2) When node $j$ is susceptible at time $t-1$, it turns into infected with probability $P_S^{j\backslash i}(t-1)-P_S^{j\backslash i}(t)$. Therefore, we have the update rule for $\phi^{j\rightarrow i}(t)$:
\begin{equation}
    \phi^{j\rightarrow i}(t) = (1-\beta_{ji})(1-\gamma_j)\phi^{j\rightarrow i}(t-1) + \left(P_S^{j\backslash i}(t-1)-P_S^{j\backslash i}(t)\right)
\end{equation}
To complete the recursion updating rules, we give the initial values as:
\begin{equation}
    \theta^{j\rightarrow i}(0)=1, \phi^{j\rightarrow i}(0)=\delta_{\sigma_j^0, I}
\end{equation}

\section{Baseline GNN}

Graph Neural Networks (GNNs) model the pair-wise interactions by implementing a trainable recurrent message passing mechanism in the graph structure, and has three main steps: \textit{Message Passing, Update and Readout}. 
After $t-1$ iterations of GNNs, let $m^u(t-1)$ as node $u$'s hidden states. In the next iteration, \textit{Message Passing} step firstly aggregates all information from  $u$'s neighborhood $\mathcal{N}_u$ as a message vector $m^{\rightarrow u}(t)$, and then node $u$ updates its hidden status to $m^u(t)$ by combining $m^u(t-1)$ and $m^{\rightarrow u}(t)$ in the step \textit{Update}. 
The updated hidden states are then fed into a task-specified \textit{Readout} function $\mathcal{R}$ for node-wise predictions in step $t$. 

% Similar to \cite{Yoon2019InferenceIP}, there are two ways to apply GNNs for MPIP. The most straightforward way is running GNNs on nodes, i.e. GNNs preserve and update the hidden status of all nodes (\textbf{node-GNN}). Imitating the updating-flow of DMP, the second method is applying GNNs on the non-backtracking\footnote{For vetexes $e_1 = (i_1\rightarrow i_2)$, $e_2=(i_3\rightarrow i_4)$ in non-backtracking line graph, there is a directed link $(e_1\rightarrow e_2)$ iff. $i_2=i_3$ and $i_1\neq i_4$.} line graph, in which GNNs preserve and update the hidden status for each directed edge (\textbf{edge-GNN}). And the hidden representations of nodes is the summation of their incoming edges hidden status.

The marginal probabilities are dependent on the nodes initial infection status $S\in \{0,1\}^{|\mathcal{V}|}$, nodes attributes $\gamma \in [0,1]^{|\mathcal{V}|} $, as well as the edge attributes $\beta \in [0,1]^{|\mathcal{E}|}$. Therefore, we first embed those attributes into vectors: 
\begin{equation}
    X_0 = \phi_n(S\oplus \gamma), \quad E_0 = \phi_e(\beta),
\end{equation}
where $\oplus$ is the concatenation operator, $\phi_n: R^2 \rightarrow R^D$, $\phi_e: R \rightarrow R^D$ and $D$ is the preset number of hidden dimension, $X_0 \in R^{|\mathcal{V}| \times D}$, $E_0 \in R^{|\mathcal{E}| \times D}$. The nonlinear functions $\phi_n, \phi_e$ are shared by all nodes and edges, respectively.

We present the details of proposed specific GNN for MPI as follows:

\begin{itemize}

\item \textbf{Step 1: Hidden status Initialization.} Unlike \cite{Yoon2019InferenceIP} initializing hidden status as zeros, we initialize the hidden states from nodes attributes: 
    \begin{equation}
        m^i(0) = \phi_{init}(X_0^i),
    \end{equation}
    where $\phi_{init}$ is a nonlinear function mapping the inputs to $R^D$.
    
\item \textbf{Step 2: Message Passing.} For node $i$ in GNN, the incoming messages $m^{\rightarrow i}(t)$ are the summations of the adjacent hidden states. To enable the messages explicitly dependent on initial conditions, before summation,  we concatenate the $E_0, X_0$ to the corresponding hidden states. Then we have:
    \begin{equation}
        m^{\rightarrow i}(t) = \phi_2 \left( \sum\limits_{j\in \mathcal{N}_i} \phi_1 \left(m^{j}(t-1) \oplus E_0^{j\rightarrow i} \right) \right),
    \end{equation}
    where $\phi_1$ and $\phi_2$ are nonlinear functions mapping the inputs to $R^D$.
    
\item \textbf{Step 3: Update.} Before updating, we concatenate the incoming messages with the target-nodes initial attributes, which is an imitation of Eq.(\ref{equ:psik}) in DMP,  to filter the incoming messages. We then update nodes hidden status with a Gated Recurrent Unit (GRU):

\begin{equation}
        m^{i}(t) = \mathrm{GRU}\left( \phi_3 \left( m^{\rightarrow i}(t) \oplus X_0^i \right), m^i(t-1)\right),
    \end{equation}
    and $\phi_3$ is a trainable nonlinear function.

\item \textbf{Step 4: Readout.} At each time step $t$ (i.e. GNN $t$th layer), the predicted marginals for node $i$ in GNN is computed by a Softmax function $ \mathcal{R}$: 
    \begin{equation}
        \label{equ:readout_gnn}
        \hat{P}^i(t) = \mathcal{R}\left( \phi_4(m^{i}(t) \oplus X_0^i ) \right)
    \end{equation}
% For edge-GNN, the node embedding $m_{i}^t$ is the summation of the incoming edges hidden status, therefore, the readout function is:
%     \begin{equation}
%         \label{equ:readout_lgnn}
%         \hat{P}^i(t) = \mathcal{R}\left( \phi_5\left( \phi_4(\sum m^{\rightarrow i}(t) ) \oplus X_i^0 \right) \right)
%     \end{equation}
where $\phi_4$ is trainable nonlinear functions. The concatenation of hidden status and $X_0^i$ aims to imitate Eq.(\ref{equ:psi}) in DMP.

\end{itemize}
Repeat \textbf{Step 2-4} until termination condition is satisfied. 

In our implementation, all the nonlinear functions $\phi_n, \phi_e, \phi_{init}, \phi_{1,2,3,4}$ are speciﬁed by MLP with ReLU as the activation function, and $\phi_{1,2,3,4}$ are shared across all time steps (GNN layers).

\section{The benefits of using the NEDMP in the within-distribution case.}
Obtaining the ground truth as the training labels for large graphs is computationally expensive. This requires a sample-efficient model. Compared to the pure data-driven baseline GNN, the proposed model NEDMP is a physics-informed hybrid model, to which the DMP module provides strong regularization. Therefore, the NEDMP is much more sample-efficient than the baseline. For two real networks, we increase the number of training samples from 3 to 52,  and the validation error of NEDMP and GNN is plotted in Figures 8 and 9. The NEDMP can achieve satisfactory error for both networks with limited training samples, while the GNN requires more training data. Therefore, using NEDMP is a better choice when obtaining the training data is expensive.

\begin{figure}[h]
  \centering
  \begin{minipage}[b]{0.45\textwidth}\label{train_diff_dolphin}
    \includegraphics[width=\textwidth]{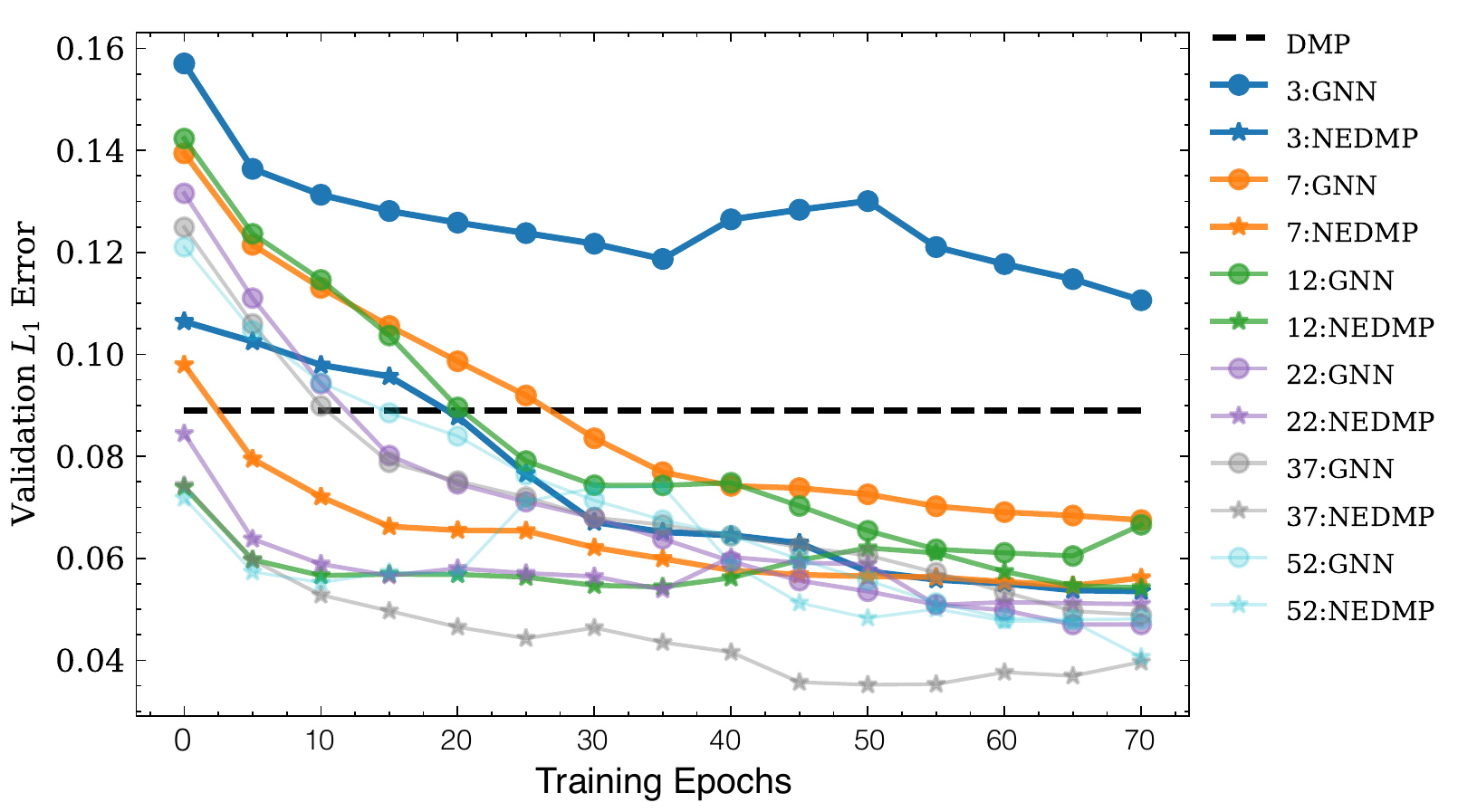}
    \caption{Val. error for network dolphins}
  \end{minipage}
  \hfill
  \begin{minipage}[b]{0.45\textwidth}\label{train_diff_norwegain}
    \includegraphics[width=\textwidth]{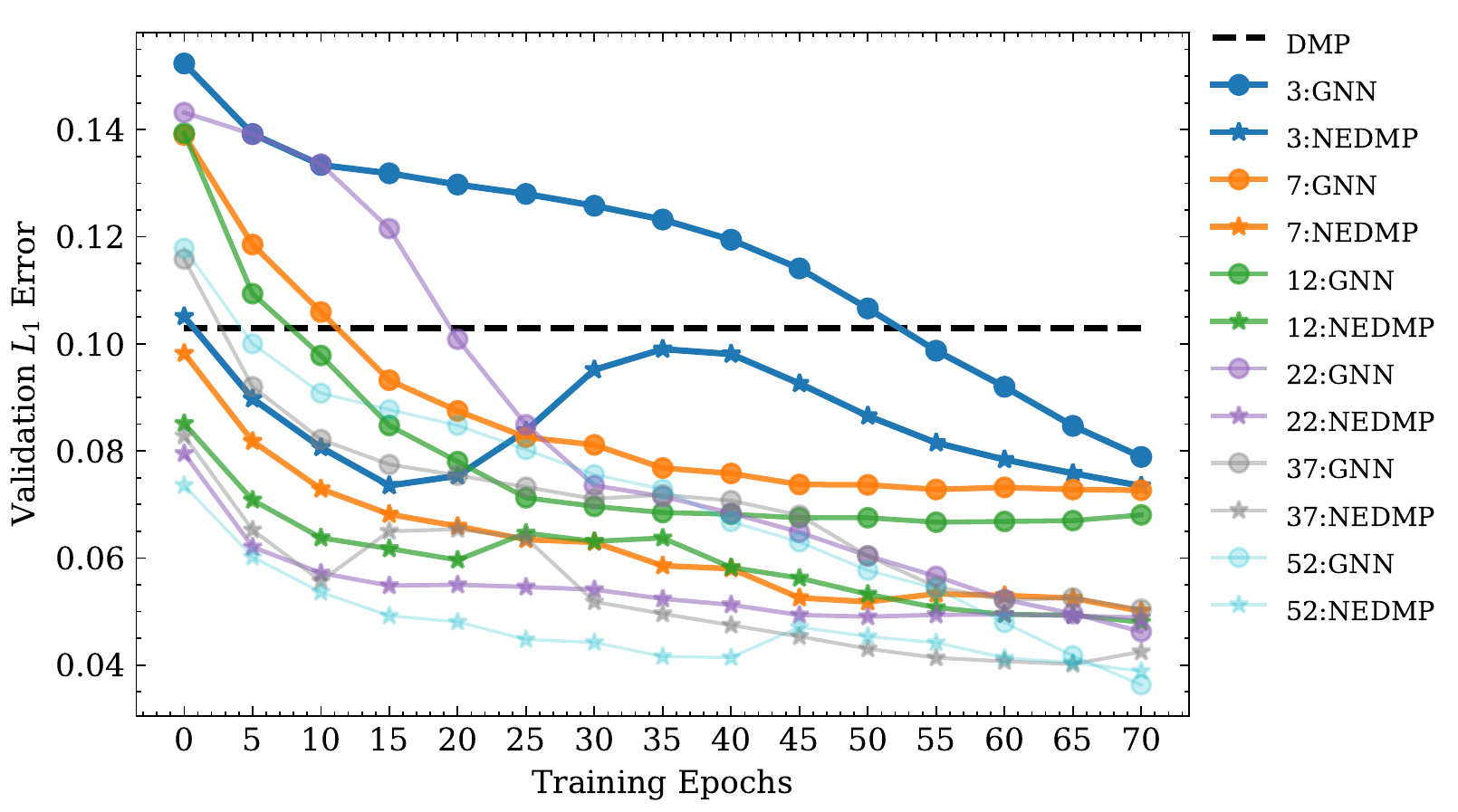}
    \caption{Val. error for network norwegain}
  \end{minipage}
\end{figure}

\section{Implementation Details}
We implement our model utilizing Pytorch\footnote{https://github.com/pytorch/} and Pytorch-Geometric. We train and test our model on Ubuntu 18.04.5 with Intel Xeon Gold 6248 CPU and NVIDIA Tesla V100 GPU.

\end{document}